# Synchronization of two memristive coupled van der Pol oscillators

M. Ignatov, M. Hansen, M. Ziegler[*)], and H. Kohlstedt[*)]

Nanoelektronik, Technische Fakultät, Christian-Albrechts-Universität Kiel,
Kiel 24143, Germany

The objective of this paper is to explore the possibility to couple two van der Pol (vdP) oscillators via a resistance-capacitance (RC) network comprising a Ag-TiO$_x$-Al memristive device. The coupling was mediated by connecting the gate terminals of two programmable unijunction transistors (PUTs) through the network. In the high resistance state (HRS) the memresistance was in the order of M$\Omega$ leading to two independent self-sustained oscillators characterized by the different frequencies f1 and f2 and no phase relation between the oscillations. After a few cycles and in dependency of the mediated pulse amplitude the memristive device switched to the low resistance state (LRS) and a frequency adaptation and phase locking was observed. The experimental results are underlined by theoretically considering a system of two coupled vdP equations. The presented neuromorphic circuitry conveys two essentials principle of interacting neuronal ensembles: synchronization and memory. The experiment may path the way to larger neuromorphic networks in which the coupling parameters can vary in time and strength and are realized by memristive devices.

[*)]Corresponding authors: maz@tf.uni-kiel.de, hko@tf.uni-kiel.de

The research on weakly coupled nonlinear oscillators goes back to Christian Huygens in 1665 when he recognized a mutual phase synchronization of two heavy mechanical pendulum clocks via a common wooden platform.[1,2] Since that time, pulse-coupled oscillators became one of the most exciting topics in science. Indeed, mutual phase synchronization of coupled nonlinear oscillators encompasses a broad range of phenomena in science and engineering. Famous examples of interacting nonlinear dynamical systems include Josephson junction arrays, fire flies, nerve cells, circadian and brain rhythms, power stations, Lasers or complex networks in general.[1-6] The underlying universality of pulse-coupled oscillator phenomena attract considerable interest and led to profound mathematical descriptions on the basis of nonlinear dynamics and graph theory.[2,6-8] Of particular interest for the present work are neuronal synchronization mechanisms of spiking neurons in the framework of information pathways within a brain.[9-12] There is persuasive experimental and theoretical evidence, that synchronized fire rates, i.e. phase synchronization between locally separated banks but reciprocally connected neuronal ensembles, are essential to explain higher brain function such as conscious awareness or decision-making.[13-17]

For the research on neuromorphic systems it might be interest to convey two important phenomena of biological computation - *synchronization* and *memory* - to an electronic circuit.[18,19] It was shown that memristive devices are able to mimic important biological concepts as Hebb´s learning rule and STDP.[20-24] In this letter we present our results on the (non-constant resistive) coupling of two self-sustained van der Pol oscillators via a memristive device. [2,25-28] In this feasibility study we demonstrate an autonomous phase and frequency locking of the two oscillators (i.e. a transition from the asynchronous to a remnant synchronous state) due to the pulse dependent memristive coupling. Our results may path the way to mimic basal coupling schemes of neuronal ensembles by memristively mediated relaxation oscillators.

The basic set-up of the experiment is illustrated in the inset of Fig. 1 (a). It consists of two self-sustained relaxation (van der Pol) oscillators, which are coupled via a memristive resistance $R_M$ *(x,t)*. Here, *x* is a state variable according to the definition of memristive devices in Ref. [29,30] and *t* the time. Experimentally, this scenario was realized using the electronic circuit sketched in Fig. 1(b). Therein, two relaxation oscillators are shown (highlighted by a red and a black frame), which are coupled through a resistor network including a single memristive device (highlighted by an orange frame).

The key device in each relaxation oscillator circuit is a programmable unijunction transistor (PUT 2N2067). [31] PUTs belong to the class of silicon controlled rectifiers and exhibit a negative differential resistance (NDR) regime in their current-voltage curve for an applied bias voltage *V* (independent variable) between the anode (A) and the cathode (C*). By applying a voltage to the gate terminal $V_G$ of the PUT, the NDR threshold voltage $V_{thG}$ can be adjusted. To get self-sustained oscillations a common circuit scheme based on a PUT oscillator has been applied [32]. It consists of a resistor-capacitor (*RC*) element at the anode side and of a voltage divider with two linear resistors ($R_B$ and $R_G$) at the gate side of the PUT (cf. Fig. 1(b)). A constant voltage $V_{BB}$ is applied to the circuit which charges the capacitor *C* via *R*. If the voltage $V_{CR}$ across *C* exceeds the threshold voltage $V_{thG}$, the PUT goes into the low resistance state and *C* is discharged to ground, which leads to self-sustained oscillations. These oscillations were picked-upped at two knots of an individual oscillator (see PUT sketch, left side of Fig. 1(b)). At knot 11 ($V_{C1}$), we observed saw-tooth type relaxation oscillations, whereas at knot 21 the voltage $V_{G1}$ exhibited short (sharp) rectangular pulses, reflecting the discharge process of $C_1$. $V_{C1}(t)$ and $V_{G1}(t)$ are equivalent in frequency and phase for an individual oscillator.

The basic idea for the realization of memristive coupled oscillators belongs to the fact, that the two oscillators are mutually coupled via the short pulses at $V_{G1}$ (knot 12) and $V_{G2}$ (knot 22), mediated via a non-constant memristance, rather than the relaxation oscillations at the knots 11 and 21. Since $V_{G1}$ varies the threshold ($V_{th2}$) of oscillator 2 and vice versa $V_{G2}$ that of oscillator 1 ($V_{th1}$), the circuit represents a so-called threshold coupled system in accordance to work of van der Pol et al..[2,27] In detail, we coupled the gate terminals $V_{G1}$ and $V_{G2}$ of two PUT-based relaxation oscillators via an *RC* network comprising a memristive device (see orange frame in Fig 1(b)). In particular, the resistor network consists of a parallel and series resistance, labelled as $R_{K1}$ and $R_{K2}$ in Fig. 1 (b), respectively. To ensure a dc potential separation between both oscillators, two capacitances $C_{K1}$ and $C_{K2}$ were part of the resistor network. In particular, $C_{K1}$ and $C_{K2}$ presents together with the resistor network a high-pass filter with a cut-off frequency defined by $f_c = (2\pi M C_K)^{-1}$. Here *M* is the resistance value of the entire memristive network. The cut-off frequency depends on memresistance $R_M$ which vary during the experiment. By using $C_{K1} = C_{K2} = 33$ nF and $R_{K2} = 47$ kΩ, $f_c$ reads 205 Hz, assuming an infinity high resistance for the initial memresistance $R_M$.

For the memristive coupling a single *Ag*-doped-*TiO$_{2-x}$*-Al memory device was used, which was fabricated on 4-inch Si wafers with 400 nm of SiO2 (thermally oxidized) in a three step photolithography process. The device stack consisting of Nb(5nm)/Ag(40 nm)/TiO$_{2-x}$(10 nm)/Al(40 nm) was defined in a mesa structure, isolated by SiO and contacted with a 500 nm Nb thick wiring. A typical I-V curve of a single *Ag*-doped *TiO$_{2-x}$* memory device is shown in Fig. 2. Therefore, an Agilent E5263A source measurement unit was employed, by sweeping the bias voltage and measuring the device current, simultaneously. A current compliance was set to precisely adjust the low resistance states (LRS) of the device, as shown in the inset of Fig. 2. For example, using a current compliance of 1 mA, the device resistance decrease from 1 MΩ to 0.5 kΩ (LRS) at a positive set voltage of $V_{set}$ = 0.6 V and vice versa for a negative reset voltage of $V_{reset}$ = -0.3 V (cf. Fig. 2). In the oscillator circuit the current compliance is realised by the serial resistance $R_{K1}$ (cf. Fig. 1(b)).

To study the phase and frequency dynamic of the entire memristive coupled two oscillator system the anode voltages $V_{A1,2} = V_{C1,2}$ (knots 11 and 12) and gate voltages $V_{G1,2}$ (knots 22 and 21) were simultaneously recorded with a Tektronix TDS 7104 oscilloscope. The obtained results are shown in Fig. 3. In the top panel of Fig. 3(a) the voltage courses $V_{A1}$ and $V_{A2}$ are plotted, while in the bottom panel $V_{G1}$ and $V_{G2}$ are represented by $V_{01} - V_{G1}$ and $V_{02} - V_{G2}$, respectively. Here $V_{01,2}$ denotes the voltage offset across $R_{G1,2}$ caused by the voltage divider $R_{G1,2}$ and $R_{B1,2}$ as well $V_{BB}$ = 20 V, which we assumed to be constant dc voltages during the charging period of $C_1$ and $C_2$. For the used circuit configuration ($R_{B1}$ = 20.8 kΩ, $R_{B2}$ = 40.5 kΩ, $R_{G1}$ = 2.37 kΩ, and $R_{G2}$ = 2.37 kΩ) we measured $V_{01}$ = 2.66 V and $V_{02}$ = 2.76 V. As a result, we found that at the beginning both self-sustained oscillators followed their own rhythm with $f_1$ = *540* Hz and $f_2$ the *410* Hz, while after ≈*36* ms oscillator 2 adapts the rhythm of oscillator 1. This is illustrated by comparing the voltage courses of oscillator 1 (black curve in Fig. 3) with that of oscillator 2 (red curve in Fig. 3). In particular, both voltages $V_{A1}$ and $V_{A2}$ rise exponentially towards *3.1* V as the capacitors $C_1$ and $C_2$ are charged through $R_1$ and $R_2$, respectively. Induced by the difference in the resistances selected for $R_1$ and $R_2$, oscillator 1 reaches the threshold voltages of PUT 1 ahead of oscillator 2. Hence, both follow their own (independent) rhythms. If the threshold voltage of one of the two PUT's is reached, a rapid decrease in resistance is caused by the negative differential resistance of the PUT. Here, the capacitor in the integration branch is rapidly discharged which leads to a rapid decrease of $V_{A1}$ and $V_{A2}$ (cf. top panel in Fig. 3 (a)). At this point the resistance of the memristive device is affected through the oscillator circuit, which produces voltage pulses $V_{G1}$ and $V_{G2}$ (cf. bottom panel in Fig. 3(a)). In particular, for each oscillation period a voltage peak is applied to the memristive device, where at some point the memristive device undergoes the transition from the HRS to the LRS. As a consequence, the coupling of the oscillators is increased and the amplitude of $V_{A2}$ is decreased, which allows oscillator 2 to synchronize with oscillator 1. We would like to emphasize that the mechanism of threshold coupling is energy efficient and therefore well suited for neuromorphic applications. [27]

In order to elaborate the mechanism of the memristive coupling in some more detail, a set of consecutive voltage pulses to set the memristive device, i.e. $V_{G1}$ - $V_{G2}$, at the desynchronized phase (AS) and at the synchronized phase (S) are compared in Fig. 3(b). While at the desynchronized phase (DS) 40 µs voltage pulses of 2.4 V are produced by the circuit, in the synchronized phase (S) width and amplitude of the pulses are decreased to 6.4 µs and 1.8 V, respectively. Moreover, while in the desynchronized phase $V_{G1}$ is 568 µs ahead of $V_{G2}$, in the synchronized phase $V_{G1}$ is only 34 µs ahead of $V_{G2}$. In particular, for a complete phase and frequency synchronization the voltage difference between the gate voltages should be zero. Hence, a phase difference is prevailed. This can be seen most clearly regarding the corresponding phase plot, depicted in Fig. 3(c). Therein, $V_{01}$ - $V_{G1}$ is plotted as function of $V_{02}$ - $V_{G2}$, where the blue and red line corresponds to, respectively, one oscillation cycle in which the memristive device is in the HRS and one cycle in which the LRS is reached. In particular, in the beginning no structured regular relation can be observed between the particular gate voltages, apart from three distinguished points in the graph, which correspond to the digital signature of the gate voltages. In contrast to this, if the memristive device has reached the LRS, some hysteretic structure (behaviour) is obtained which is typical for two oscillators with equal frequency and a constant phase shift [2]. In other words, the transition of the memristive device resistance from the initial HRS to the LRS leads to frequency synchronisation with a constant phase difference.

To understand the underlying principles of the oscillator system in some more detail, the system has been modelled by two mutually coupled *van der Pol* oscillators with the following set of second-order dimensionless nonlinear equations:

$$\frac{d^2x_1}{dt^2} - \alpha_1(1-x_1^2)\frac{dx_1}{dt} - \beta_1 x_1 \frac{(x_1-\gamma_1)^2}{\gamma_1^2} = m(x,t)(x_2-x_1)$$
$$\frac{d^2x_2}{dt^2} - \alpha_2(1-x_2^2)\frac{dx_2}{dt} - \beta_2 x_2 \frac{(x_2-\gamma_2)^2}{\gamma_2^2} = m(x,t)(x_1-x_2)$$
(1)

Here $\alpha_{1,2}$, $\beta_{1,2}$, and $\gamma_{1,2}$ are positive constants detuning the frequencies and damping behaviour of the uncoupled oscillators, while $m(x,t)$ is the positive coupling strength for the coupled system with the memory state variable $x$. Guided by the experiment we assumed a binary changes from a weak (or non-) coupling regime $m_0$ ($R_M$ in HRS) to a strong coupling $m_1$ regime ($R_M$ in LRS) defined by $m(t) = m_0 \Theta(t-t_s) + m_1 \Theta(t_s-t)$. Here, $\Theta(...)$ is the *Heaviside* step function and $t_s$ the time of the resistance switching, which is used as an independent fit parameter. Moreover, a slight modification of the basic form of the *van der Pol* oscillator has been made to take the digital spike line shape of the experimental oscillator into account (cf. $V_{G1}$ and $V_{G2}$ in Fig. 3 (a)). Therefore, a nonlinear cubic term $[x_{1,2}(x_{1,2}-\gamma_{1,2})^2/\gamma_{1,2}^2]$ has been added close to the investigated system in [33], where $\gamma_{1,2}$ can be used to define the number of spikes per unit time of the respective oscillator in the uncoupled phase. The obtained oscillation dynamics of $x_1$ and $x_2$ are depicted in Fig. 4(a). In agreement with our experimental findings, an uncoupled oscillation of $x_1$ and $x_2$ is observed if the coupling strength is low ($m_0$ = 0.01), while for an increased coupling ($m_1$ = 0.1) the system synchronizes, as shown in Fig. 4 (a). To analyze the synchronization behaviour in some more detail $x_1$ is plotted versus $x_2$ in Fig. 4(b). As already done for the experimental data, shown in Fig. 3(c), one low coupling oscillation cycle ($m_0$ = 0.01) is highlighted in blue and one high coupling cycle ($m_1$ = 0.1) is plotted in red. We found that our modified *van der Pol*-model enables to confirm the main experimental findings, such as a independent oscillation for a low coupling strength ($m = m_0$), which is indicated by an "*L*"-line shape in the phase plot and a closed hysteretic loop for $m_1$, i.e. frequency synchronization and phase locking.

A more detailed understanding of the frequency synchronization can be obtained by regarding the frequency mismatch between the uncoupled system $\Delta f = f_1 - f_2$ and the coupled system $\Delta F = F_1 - F_2$ in dependence of the coupling strength, as depicted in Fig. 4(c). The frequency detuning was measured for three different coupling resistances: An infinitely high resistance (blue curve), 10 kΩ (orange curve), and 1 kΩ (red curve), while for the simulation *m* has been chosen to 0.01 (blue curve), 0.05 (orange curve) and 0.1 (red curve). Further, the frequency of oscillator 1 has been hold constant at 414 Hz ($f_1 = f_0$), while $f_2$ has been swept in between 310 Hz and 640 Hz, for the three different couplings. For a better illustration and comparison between simulation and measurement the obtained frequency mismatches are normalized by $f_0$. We found that for low coupling strengths (blue curves in Fig. 4(c)) a large frequency mismatch prevent synchronization, while for stronger couplings the width of the synchronization regime is drastically increased (cf. orange and red curves in Fig. 4(c)). Noticeable is the asymmetry in $\Delta f$ between the positive and negative synchronization regime (see orange curves in Fig. 4(c)), which is even more pronounced in the experiment compared to the simulation. This asymmetry results from the fact that the oscillator system will be synchronized always towards the oscillator with the higher frequency.

In conclusion, frequency synchronization and phase locking of two memristively coupled self-sustained relaxation oscillators has been shown. Therefore, two PUT relaxation oscillators are coupled via a resistor network consisting of an *Ag*-doped-*TiO$_{2-x}$*-Al memristive device. A theoretical model of two modified van der Pol oscillators has been employed to support the experimental findings and conveys two essentials principles of the presented electrical circuit: synchronization and memory. In this regard, we strongly belief that our investigation path the way to a new class of dynamical networks in which the coupling parameters can vary in time and strength and are realized with memristive devices exhibiting binary or analogue switching *I-V* characteristics.

This work was also supported by the Deutsche Forschungsgemeinschaft (DFG) via the Research-Group grant FOR 2093.

**Figure captions:**

**Fig. 1:** (a) Illustration of two mutually coupled self-sustained *van der Pol*-oscillators. The memory is introduced through the memristive coupling $R_M(x,t)$, where $x$ is the state variable of the memory process. The frequencies $f_1$ and $f_2$ are the frequencies of the uncoupled oscillators. (b) Circuit scheme used to implement a memristive coupling of two van der Pol-oscillator experimentally: The circuit consist of two relaxation oscillators (highlighted by the two dashed frames) which are functionally based on *programmable unijunction transistors* (PUTs). $A_{1,2}$, $C_{1,2}$, and $G_{1,2}$ are the anode, cathode and gate terminal of the two PUTs (1 and 2), respectively. Through $V_{BB}$ a constant voltage is applied, which ensures the self-sustain oscillation. The individual oscillator are mutually coupled via their gate terminal through a memristive resistor network $M(x,t)$ (highlighted by an orange frame). The coupling network consists of a single memristive device $R_M$ in series with a resistor $R_{K1}$, a parallel resistor $R_{K2}$ as well as two capacitances $C_{K1,2}$. The basic idea of the circuit is to vary the gate potentials of the PUTs in dependence of the resistance state of the memristive device, as explained in the text.

**Fig. 2:** (a) $|I|$-V curve of an Ag-doped $TiO_{2-x}$ based memristive device with two different settings of the current compliance $I_{CC}$. Inset: The low resistance values $R_{LRS}$ as a function of the applied compliance currents.

**Fig. 3:** Circuit performance: (a) voltage traces recorded for $V_{A1}$ and $V_{G1}$ (black lines) and $V_{A2}$ and $V_{G2}$ (red lines). (b) Difference in the gate potentials $V_{G1}$-$V_{G2}$ before (labelled as desynchronized phase *DS*) and after resistance switching (labelled as synchronized phase *S*). (c) phase plot of the system. One cycle of the DS phase (HRS) and S phase (LRS) have been highlighted in blue and red, respectively. (Device parameters: $R_{B1}$ = 20.8 kΩ, $R_{B2}$ = 40.5 kΩ, $R_{G1}$ = 2.37 kΩ, $R_{G2}$ = 2.37 kΩ, $R_{K2}$, = 47 kΩ, $V_{BB}$ = 20 V, and $C_{K1}$ = $C_{K2}$ = 33 nF )

**Fig. 4:** Simulation results of two memristively coupled van der Pol oscillators: (a) System dynamic for oscillator 1 (black curve in the upper panel) and oscillator 2 (red curve in the upper panel) under a changed coupling strength m (lower panel). (b) Corresponding phase plot of the oscillators. One cycle with *m* = 0.01 and *m* = 0.1 are plotted in blue and red, respectively. (c) Frequency detuning for different coupling strength for the van der Pol model (upper graph) and electrical circuit (lower graph). While *Δf* is the frequency mismatch of the uncoupled oscillators, *ΔF* denotes the frequency mismatch of the mutual coupled oscillators. (Simulation parameters: $\alpha_1 = 3.5$, $\alpha_2 = 4.8$, $\beta_{1,2} = 0.1$, $\gamma_{1,2} = 3.0$, $m_0 = 0.01$, $m_1 = 0.1$).

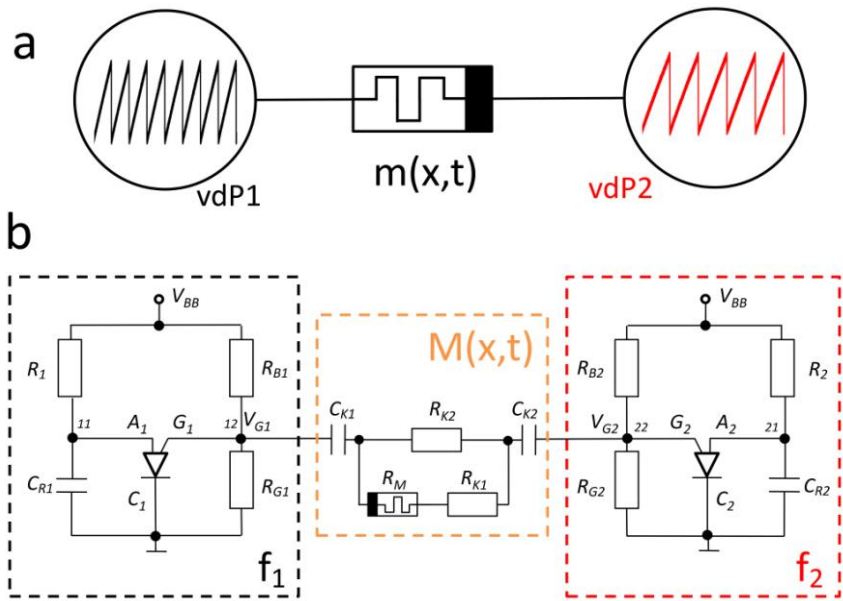

Fig. 1

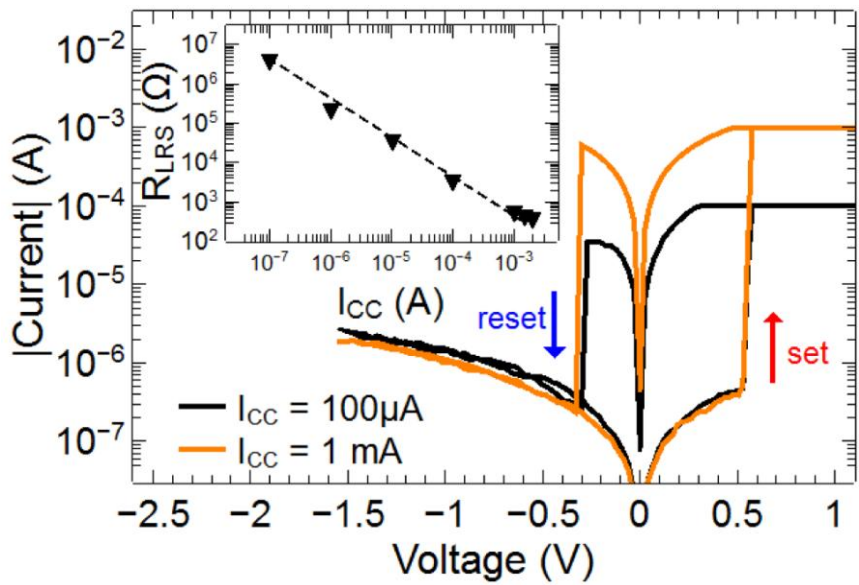

Fig. 2

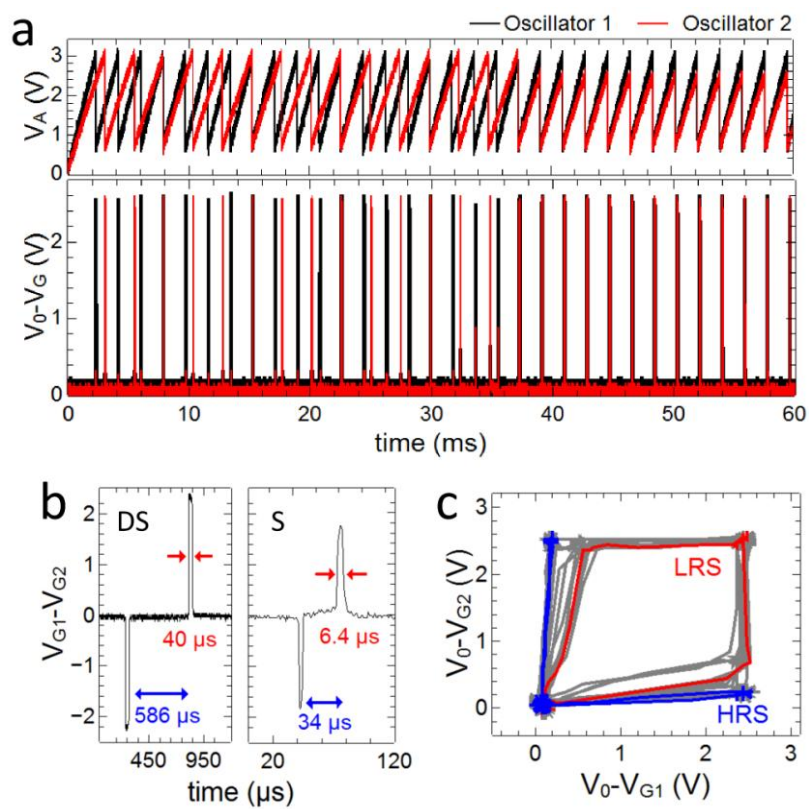

Fig. 3

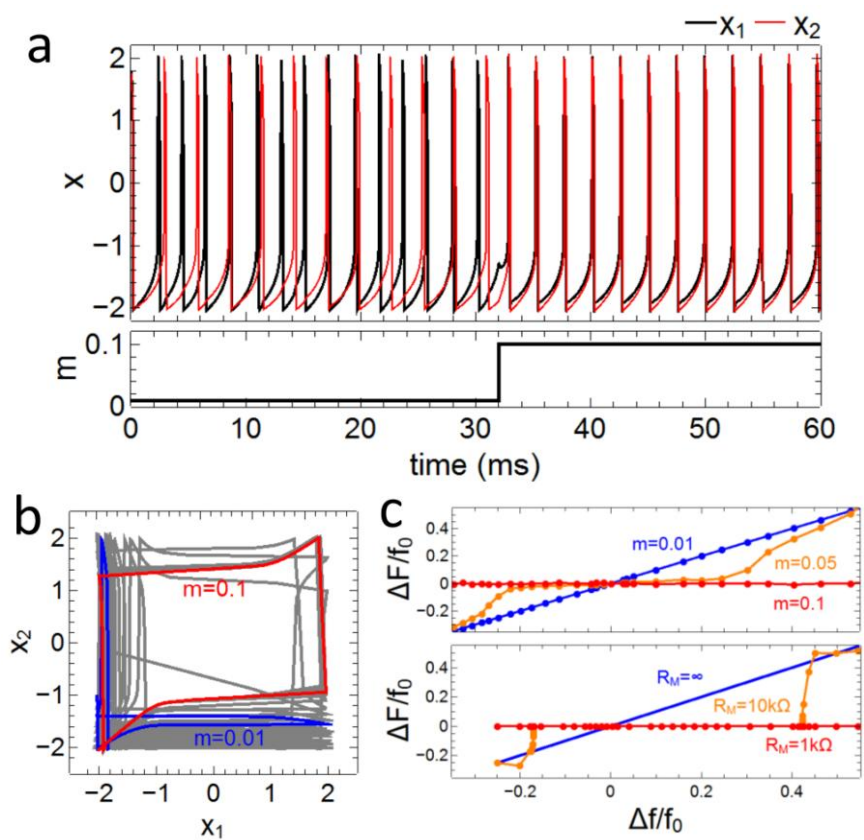

Fig. 4